# Mind the gap: how multiracial individuals get left behind when we talk about race, ethnicity, and ancestry in genomic research

Daphne O. Martschenko[1], Hannah Wand[2], Jennifer L. Young[1,3], Genevieve L. Wojcik[4]

1. Center for Biomedical Ethics, Stanford University School of Medicine, Stanford CA
2. Department of Cardiology, Stanford University School of Medicine, Stanford CA
3. Center for Genetic Medicine, Northwestern University, Feinberg School of Medicine, Chicago IL
4. Department of Epidemiology, Johns Hopkins Bloomberg School of Public Health, Baltimore MD

## Abstract

It is widely acknowledged that there is a diversity problem in genomics stemming from the vast underrepresentation of non-European genetic ancestry populations. While many challenges exist to address this gap, a major complicating factor is the misalignment between (1) how society defines and labels individuals; (2) how populations are defined for research; and (3) how research findings are translated to benefit human health. Recent conversations to address the lack of clarity in terminology in genomics have largely focused on ontologies that acknowledge the difference between genetic ancestry and race. Yet, these ontological frameworks for ancestry often follow the subjective discretization of people, normalized by historical racial categories; this perpetuates exclusion at the expense of inclusion. In order to make the benefits of genomics research accessible to all, standards around race, ethnicity, and genetic ancestry must deliberately and explicitly address multiracial, genetically admixed individuals who make salient the limitations of discrete categorization in genomics and society. Starting with the need to clarify terminology, we outline current practices in genomic research and translation that fail those who are 'binned' for failing to fit into a specific bin. We conclude by offering concrete solutions for future research in order to share the benefits of genomics research with the full human population.

# Introduction

The majority of human genomic research is built upon an overrepresentation of European genetic ancestry populations,[1] concentrating potential benefits to a narrow subset of the global population and potentially exacerbating health disparities.[2] To address these issues, the genomics community has committed to enhancing diversity in genomics research. However, a major complication to these efforts is the disconnect between how the genomics community defines populations, and how society perceives and uses population descriptors, often conflating race, ethnicity and ancestry.[3] While recent ontologies in the field seek to improve the specificity of terms used to define populations to prevent this misunderstanding, they continue to discretize the continuous spectrum of human diversity, enabling the incorrect understanding of race as biological to persist.[4,5]

Additionally, these frameworks largely ignore the issue of how to utilize data from people of multiracial backgrounds who cannot be discretized, leading to their continued exclusion from biomedical research and subsequent downstream potential benefits.[6] Rather than following the status quo, the genomics community must think creatively about inclusivity; bins of people may be useful and straightforward during the early stages of research, but they inherently limit the translational benefits of research to real populations. Until the field of human genomics adopts frameworks that are inclusive of everyone, genomics research and medicine will remain discriminatory and unjust.

We argue that any novel framework to establish standards around race, ethnicity, and genetic ancestry must deliberately and explicitly address multiracial individuals – a rapidly growing population that tends to have blended genetic ancestry and more flexible social identities.[7] Developing frameworks adapted to those who constitute a diverse constellation of racial, ethnic, and genetic ancestry backgrounds would necessarily disrupt the research community's and society's reliance on discrete categorization.

As a multidisciplinary team of multiracial researchers, with diverse professional expertise across the fields of bioethics, population genetics, genetic epidemiology, genetic counseling, translational research, and family science (See **Supplementary Note 1**), we outline current practices in genomics that fail multiracial individuals in both research and research translation. We begin this Commentary with a discussion of terminology and the need to be clear and intentional in how we define populations. We then provide recommendations for agenda-setting, research conduct, research translation, and research communication to ensure that the benefits of genomic research and precision health are accessible to everyone.

## What Is 'Multiracial'? A Note on Terminology

What it means to be multiracial is complex.[8] As **Box 1** illustrates, there are many different terms researchers use to describe individuals who cannot be neatly classified into a homogenous, often continental-level, group. For the purposes of this Commentary, we use the terms 'multiracial' to describe individuals with backgrounds from multiple races. However, we recognize that there is a lack of consensus within the research community on how to define diversity measures such as race, ethnicity, and ancestry.[9]

> **Box 1. The challenge of defining multiracial individuals.**
>
> A challenge lies in the description of multiracial individuals with both precision and inclusivity. The EBI-NHGRI GWAS Catalog currently collapses genetic ancestry, geography, nationality, and race when determining the 'ancestry' of GWAS participants. Participants can only be assigned to a single category, such as "Other admixed ancestry" or the more specific "African American or Afro-Caribbean" or "Hispanic or Latin American".[11] However, individuals may also fall into "Other" or "Not Reported" depending on the individual study's definitions. The All of Us research program lists "More than one race/ethnicity" as a category with 6.6% of their participants. However, the collapsing of race and ethnicity within a single question makes this number uninterpretable on its own.[12]
>
> While we have chosen to use "multiracial" throughout the commentary to ensure inclusivity grounded in social contexts, not genetic, other terms are used in the literature. This is an illustrative list intended to demonstrate heterogeneity and complexity amongst individuals who are not monoracial. It is neither comprehensive nor weighted in importance or usage.
>
> **Broad Terms**
> Many of terms that are often used for multiracial individuals are broad and non-specific. Additionally, they may refer to concepts that are unrelated to the construct relevant to a specific genomic study.
> - **Admixed**: refers to genetic ancestry independent of racial or ethnic identity and can include both multiracial and single race individuals (e.g., African American, Hispanic/Latino groups)
> - **Multiracial, biracial, mixed race**: referring to race, with biracial being limited to only two racial categories and mixed race being derogatory in certain contexts
> - **Multicultural**: referring only to culture and not to race, ethnicity, or genetic ancestry
> - **Other**: a catch-all term that lacks precision and promotes 'otherization' of individuals
>
> **Specific Terms**
> Other terms are too specific to refer to multiracial individuals at large, and many originate from specific violent histories or a particular cultural context. In general, these terms are neither broadly accepted nor well-understood amongst the individuals who may be referred to as such. Two examples are 'Mulatto' and 'Hapa.' Mulatto has origins in enslavement and is considered outdated and offensive; it refers to individuals of African and European ancestry. Hapa is used to refer to individuals of Asian and European ancestry. It is misappropriated from the Hawaiian term 'hapa haole' or a part White and part Hawaiian individual, and is thus problematic when used broadly.
>
> In short, terminology used in genomic studies, or any study in general, should be responsive to both the cultural contexts and the scope of ancestries included, genetic or non-genetic. A lack of precision and inclusiveness in terminology can hinder the characterization and translation of genomic findings, limiting downstream benefits.

Despite these caveats, being mindful of and transparent about how we define terms will help to clarify the relationship between definitions and (1) how we conduct research, including who we choose to include and not include in our studies; and (2) how we translate and communicate research findings to support the health of everyone. Decisions about how to classify people are rooted in a history informed by social, economic, and political influences.[10] A failure to acknowledge these processes will perpetuate problematic uses of language.

Furthermore, we recognize that the term 'multiracial' indicates an ascribed category that may come to represent yet another way to categorize people into discrete groups and reinforce rather than disrupt dominant approaches to classification. However, to have a concrete discussion about inequities and care gaps in genomics, conceptualizing an illustrative 'group' for discussion proves useful. Any use of the term 'multiracial' in this Commentary is not intended to mask the diversity in cultural, ethnic, and lived experiences among those who do *not* identify (by self or others) as one race, ethnicity, or ancestry. It is intended as a broadly inclusive term, with individuals holding varied conceptions of their identity.

Additionally, use of the term 'multiracial' in genomics should be mindful of the dangerous slippage between race and genetic ancestry. 'Multiracial' should not be used as a proxy for genetic admixture and vice versa. While there is often a correlation between individuals who are admixed (recent ancestry from two or more populations) and also identify as multiracial, race is ascribed by oneself or others based on physical characteristics. As such, some individuals may not identify or be recognized as multiracial, instead identifying with a monoracial group (e.g., 'Asian,' 'Black', etc.) based on salient physical features. In short, 'multiracial' cannot be assumed as a distinct and non-overlapping group and discretizing individuals may not be congruent with current lived experiences. Despite these limitations, we will use 'multiracial' throughout this Commentary as an inclusive term, while acknowledging its strengths and limitations.

## Agenda Setting & Justice

Ensuring that all members of society benefit equitably from scientific research is central to achieving justice, a context-dependent concept broadly entailing treating people fairly by giving each person their due. Currently there is disproportionate funding of research for individuals and conditions that affect a relatively small and privileged subset of the global population; this is problematic considering much of research funding is allocated from public, taxpayer money.[11] In an attempt to share the benefits of genomic research with the full human population, genomics researchers and funding agencies are increasing efforts to diversify biobanks and include community stakeholders' views in setting research agendas and assessing community needs.[12] Building partnerships between researchers and community members builds trust as researchers are more likely to respect community values, anticipate potential risks, and optimize potential benefits.[13,14]

Community engagement is central to achieving justice in genomics research. However, it is only successful if one can define and reach a community. To that end, it is particularly difficult to engage multiracial communities due to the small but growing numbers of multiracial individuals who may or may not identify as a coherent group. Furthermore, as multiracial individuals are engaged to discuss their research needs or concerns, researchers should be prepared to adjust expectations around consensus and shared identity given the heterogeneity of this group. It cannot be expected that models of interacting with homogeneous groups will extend to this population, and there are no current frameworks to guide researchers on how to define, recruit, and engage a community of multiracial people.

While it is necessary to incorporate multiracial perspectives into the design and conduct of genomics research from a justice perspective, additional challenges remain. A novel framework is needed to both recruit and engage a multiracial community members and multiracial researchers to aid in study design and help foster trust with participants. Therefore, both funders and researchers need to (1) cultivate an ethos of diversity in the workplace that appreciates the perspectives of multiracial researchers[15]; and (2) prioritize building dynamic, bi-directional communication between multiracial community members and researchers from the beginning of the research process. Critically, achieving these aims requires trust between community members and researchers, and the existence of trustworthy institutions. One step towards building trust is to enhance transparency about who researchers are as individuals and what motivates them. Frequently utilized in ethnographic, qualitative research, but underutilized in quantitative research such as genomics, positionality statements explicate researchers' existing biases by outlining the position that a researcher has adopted within a given research study (See **Supplementary Note 1**). Recognizing the role and relevance of researchers' positionalities for agenda setting, research conduct, and research translation, might humanize the field of human genetics while instilling the cultural humility required for successful community engagement.

Funders can help set these priorities within the research community and affect institutional culture through funding calls to consider the unique and multi-faceted experiences of multiracial individuals in genomics. Revisiting funding priorities to account for multiracial individuals can encourage more inclusive and person-centered agenda setting and research designs that prioritize justice, trust, and trustworthiness, and maximize the benefits of genomics research for all.

## How genetic and genomic research currently account for multiracial individuals

Once research is funded, it is often expected to translate into meaningful goods for public knowledge, health or consumption – thus creating a cycle between public agenda setting, research, translation and public benefit. Every phase of the genomic research pipeline, including the recruitment of participants, the analysis of curated data, and the interpretation of results, has been shaped by the assumed default of homogeneous European ancestry populations. The expansion to other populations has followed the same model, in which data is often stratified by a combination of race or ethnicity (often self-identified) and/or genetic ancestry.[16]

The recruitment and representation of multiracial individuals in genomic studies is limited and difficult to quantify in large-scale genomic studies.[17] Beyond participation, there remain steps along the data processing and analytical pipeline in which those who do not fit within a single box, whether delineated by race/ethnicity or genetic ancestry, are filtered out. Multiracial individuals are often removed during preliminary quality control (QC) steps which stratify based on race, ethnicity, or a homogeneous genetic ancestry grouping.[5,18] In the absence of these groupings, investigators often assign a label – such as by principal components using recent machine learning methods. If an individual's determined genetic ancestry does not fit cleanly within a cluster, they are dropped from further analysis.[19]

Once past the constraints of data processing and QC, multiracial individuals often also find themselves excluded from the discovery stage. It is common to stratify large-scale genomic studies, such as genome-wide association studies (GWAS), by either genetic ancestry with often arbitrary cut-offs, or by race/ethnicity.[18] However, newer methods now allow the pooling of individuals from multiple genetic backgrounds, leveraging shared ancestry across groups.[22] The discretization of genomic data along racial and/or ethnic lines can also hinder the benefits for precision health.[16] For example, polygenic scores (PGS) often rely on previous GWAS which compounds the exclusion of multiracial individuals by relying on previous groupings and carrying forward those definitions when composing a reference distribution of genetic risk.[2]

Indeed, throughout genomic research, there are important outstanding questions regarding 'lumping and splitting' individuals, and the extent to which individuals should be grouped for data to be meaningful. Taken together, these practices within genomic research compound and result in a dearth of results applicable to multiracial individuals, restricting any downstream benefits. While recently the field of genetics has renewed efforts for the development of statistical methods inclusive to diverse populations, it is essential that these be done, when possible, with consideration towards these limitations with the goal of analytical frameworks that acknowledge human diversity as a complex spectrum not adequately accounted for in discrete bins.

## How the use of genomics in clinical care and population health discounts multiracial individuals

Multiracial individuals face potential health disparities in clinical practice at two levels: first at the level of genetic test availability and performance, and second at the level of accessing genetic services. The availability and performance of certain genetic tests (e.g., next generation sequencing, PGS, etc.) are inherently limited to the populations represented in the research. As such, the baseline comparison for human genetic variation is biased towards an understanding of variation in European ancestries, meaning that non-White patients, including multiracial individuals, have a higher rate of 'uncertain' results due to the discovery of variants that are not represented in population databases.[20] Despite efforts to diversify biobanks and provide greater, more equitable access to clinical genetic services by externally validating PGS across major ancestral groups, multiracial individuals specifically remain neglected as they are often excluded from PGS validation studies, automatically disqualifying them from the benefits of downstream clinical testing. Without consideration of multiracial individuals in genomic studies, the translation of research to clinical genomics is likely to lead to inappropriate care and management.

Additionally, enhancing genetic test performance for multiracial individuals will not necessarily translate into access. There are persistent barriers in access to genetic services that stem from how race, rather than genetic ancestry, is treated in medicine. Genetic ancestry largely is not measured or used by clinicians or public health researchers. Health disparities related to access, and subsequent clinical outcomes, are evaluated through public health research at a health systems level by looking for differences in care along axes of social determinants of health, including race.[16] Race-based disparities can only be measured if race is accurately documented and available, yet the nature and extent of disparities for multiracial individuals are currently unclear due to difficulties in documenting multiracial as a demographic value.

Race-based disparities in medicine arise from implicit provider bias and racial profiling[21], as well as systemic barriers from correlated socioeconomic disparities[22] (e.g., insurance coverage). In clinical genetics, differential treatment of patients by race occurs in genetic test access, genetic counseling access, provider access, and access to downstream services following genetic testing. Specifically, racial and ethnic minorities are offered fewer genetic services and have less coverage of services. How race-based disparities in clinical genetics are experienced by multiracial individuals in particular is unclear. For instance, little is known about how multiracial individuals are subjected to structural barriers – for example when one parent belongs to a more privileged racial group (typically White). Multiracial individuals are likely to have varied experiences in the perception of care or access services based on their specific backgrounds due to heterogeneity in socially-assigned race and self-identity.[23] It should not be assumed that their experiences will be a 'sum of parts' of their various backgrounds.

Given the dearth of research on the experiences of multiracial individuals in clinical genetics, attention should be devoted to understanding how: (1) multiracial individuals are racially profiled and treated by clinical geneticists, genetic counselors, and physicians, and (2) how other social determinants of health correlate with being multiracial. While population health research in these areas is necessary to document and understand disparities faced by multiracial individuals, it signifies the start of the conversation, not a solution. The issues presented here reinforce the need for our medical system to move away from binaries to more inclusive care along a spectrum of personal experiences and genetic backgrounds. Before genomics can be integrated into public health programs, firm scientific evidence assessing the role of social determinants of health is necessary.

## Moving forward to close the gap

Any conceptual or empirical framework that is unable to accommodate the 'fringe' cases that multiracial individuals represent will perpetuate our overreliance on discrete categories. If we are going to reimagine the field of human genetics, it needs to be done in an inclusive manner that accommodates changing demographics and includes multiracial individuals in health equity efforts. **Table 1** draws upon the prior sections of this Commentary to summarize our recommendations for achieving these objectives during each step of the research process.

**Table 1. Suggestions to close the gap and include multiracial individuals in genomic research**

| Stage of Scientific Process | | Suggestions to Close the Gap in Service of Public Health |
|---|---|---|
| Agenda-Setting | Engage Multiracial Individuals in Decision Making & Health Equity Initiatives | • Engage diverse array of multiracial stakeholders, including both participants and researchers<br>• Require clear definitions of diversity or health disparity, inclusive to multiracial individuals in funding calls |
| Research Conduct | | • Recruit more multiracial participants<br>• Avoid binning participants into discrete racial/ethnic/ancestral categories<br>• If possible, pool individuals from multiple genetic backgrounds, leverage shared ancestry across groups |
| Research Translation | | • Include multiracial groups in health equity efforts<br>• Expand research frameworks to include the experiences of multiracial individuals in healthcare systems<br>• Prioritize person-centered clinical genetics practice |
| Research Communication | | • Establish and follow standard guidelines on reporting race/ethnicity/ancestry, inclusive to multiracial individuals<br>• Avoid grouping multiracial individuals into single monolithic categories<br>• Disclose authors' positionality and biases |

Multiracial individuals bring into question the purpose and utility of continental groupings in genomics by illustrating the complexity of human history in a field that has long sought to reduce such complexity. Therefore, as we reimagine standards on the reporting and treatment of race, ethnicity, and genetic ancestry, it is imperative to develop a framework flexible to multiracial individuals. This need for flexibility is crucial as we develop novel genomic tests (e.g. PGS) to aid in clinical care. We must move away from the possible reification of race-based medicine and towards a full accounting of the complexities of genetic ancestry and environmental contributors to health when integrating genomics into clinical research and care.

Accounting for those who are often left out of genomic studies will enhance access and inclusivity, something genomics has thus far struggled to provide. Given the burdened history behind the field of human genetics, a commitment to operating with respect and intentionality will enable us to celebrate our differences and support the flourishing of the full human population.

# Supplementary Note 1: Author Positionality Statements

**DOM** *(she/her)*: I identify as a biracial African American. My parents both immigrated to the United States – my father as a child from the Ukraine and my mother as an adult from Nigeria. I was born in England and spent portions of my childhood living in various Eastern European countries, ever aware of the curiosity my biracial family sparked in passersby. My multicultural upbringing and challenges with my racial identity motivate my work investigating the social and ethical implications of genetic/genomics and in particular the fraught and violent relationship between race, genetics, and human behavior.

**HW** *(she/her)*: I identify as a biracial Asian American. My mother grew up as an orphan in Korea, and immigrated to the US as an adult. My father's family immigrated from Poland. They met in college and built a family of choice with other immigrants. Both of my parents had mixed feelings towards their national/cultural origins: my mom holds a lot of resentment for how Korea treated orphans as a social class, and my father is an atheist from a Polish Catholic community. This led to them having a scattered sense of cultural identity, and where they lapsed, I adopted values and traditions from my other "family" members. This blended upbringing made me interested in the complexity of intersecting genetics with self-identity, family, generational health, and culture. It's driven my clinical work as a genetic counselor, and research in public health genetics where societal interactions with genetics create complex challenges.

**GLW** *(she/her)*: I am unsure as to how I identify, as my experience as a biracial individual in the United States has largely been defined by what I am not, instead of what I am. My mother immigrated from Taiwan and my father's parents from France and Poland. My research interests in genetic epidemiology for diverse, and specifically admixed populations, have been partially motivated by my background to ensure that discoveries will also benefit my loved ones, whether family or friends, with increased urgency for my multiracial children.

**JLY** *(she/her)*: I identify as a biracial Asian American. My parents both immigrated to the United States, one from China and one from England. Growing up in a very international neighborhood in midwestern United States exposed me to a range of different cultures, but people who identified as multiracial were still a significant minority in my hometown. Issues of racial identity and cross-cultural families has been at the center of my family systems research as well as my clinical work as a family therapist.